\documentclass[smallextended]{svjour3}       % onecolumn (second format)
\smartqed  % flush right qed marks, e.g. at end of proof
\usepackage{graphicx}
\usepackage{cite}
\usepackage{xcolor}
\usepackage{url}
\usepackage{lipsum,booktabs}
\begin{document}

\title{Plasmonic sensors based on funneling light through nanophotonic structures}



\author{
Mahmoud H. Elshorbagy, Alexander Cuadrado,  and Javier Alda
}


\institute{M. Elshorbagy,  and J. Alda \at Applied Optics Complutense Group. University Complutense of Madrid. Faculty of Optics and Optometry. Av. Arcos de Jalon, 118. 28037 Madrid, Spain.
               \\
              Tel.: +34.91.394-6874\\
              Fax: +34.91.394-6880\\
              \email{mahmouha@ucm.es}, and
              \email{javier.alda@ucm.es}           \\
A. Cuadrado  \at Escuela Superior de Ciencias Experimentales y Tecnolog\'{\i}a. Universidad Rey Juan Carlos. M\'{o}stoles, 28033 Madrid.\\
\email{alexander.cuadrado@urjc.es} \\
           M. Elshorbagy is also with \at
           Physics Department. Faculty of Science. Minia University.  61519 El-Minya, Egypt.\\    
           Corresponding author: Javier Alda
}

\date{This is a pre-print of an article published in {\em Plasmonics}. The final authenticated version is available online at: \url{
https://doi.org/10.1007/s11468-019-01105-6}.}

\maketitle


\begin{abstract}

We present a refractometric sensor realized as a stack of metallic gratings with subwavelength features and embedded within a low-index dielectric medium. 
Light is strongly confined through funneling mechanisms and excites resonances that sense the analyte medium. 
Two terminations of the structure are compared. One of them has a dielectric medium in contact with the analyte and exploits the selective spectral transmission of the structure. 
The other design has a metallic continuous layer  that generates surface plasmon resonances at the metal/analyte interface. 
Both designs respond with narrow spectral features that are sensible to the change in the refractive index of the analyte and can be used for sensing biomedical samples.

\keywords{Nanophotonics \and Plasmonics \and Optical Sensors  \and Refractometry }

\end{abstract}

%%%%%%%%%%%%%%%%%%%%%%%%%%  body  %%%%%%%%%%%%%%%%%%%%%%%%%%

\section{Introduction}

Plasmonic sensors are used as refractometers for the identification and characterization of chemicals \cite{JORGENSON1993213}, gases \cite{Bingham2010}, and biomaterials \cite{Kabashin2009}.  
The excitation of surface plasmon resonances at the metal/dielectric interface is reinforced and favored by the use of photonic nanostructures  \cite{brolo_lamgmuir_04} 
and nanoparticles \cite{li2015, du2018}. 
Some of these designs allow normal incidence conditions that help its integration with optical fibers and avoids angular interrogation \cite{Elshorbagy2017,Elshorbagy2019}.
In optical sensing is customary to use two characteristic parameters: sensitivity ($S_B$) and figure of merit (FOM).
They are defined for spectrally interrogated devices as follow 
\cite{homola_saa_97,vangent_appopt_90, Elshorbagy2017prism}:
\begin{equation}
S_{B}=\frac{\Delta \lambda_{\rm res}}{\Delta n_a}
\label{eq:sens}
,\end{equation}
\begin{equation}
{\rm FOM}=\frac{S_{B}}{\rm FWHM}
,\label{eq:FOM}
\end{equation}
where $\Delta \lambda_{\rm res}$ is the wavelength shift of the reflectance, transmission or absorption dip (or peak) due to a change $\Delta n_a$ in the refractive index of the analyte. FWHM is the full width at half maximum of the spectral response lineshape.
Eq. (\ref{eq:sens}) can also be seen as the derivative of the location of the spectral feature with respect to the index of refraction of the analyte. When this dependence, $\lambda_{\rm res}(n_a)$, is linear over a given range in $n_a$, sensitivity remains constant within this range. These parameters account for the performance of the systems when comparing  sensing strategies, and quantify how the system detects changes in the sensed property (for example,  the index of refraction of the analyte). 
Currently, there is a great interest in raising the values of these performance parameters and provide new designs and strategies to ease fabrication and operation of the sensor.
Actualy, we have found in the literature an enhanced sensitivity of 2459.3 nm/RIU for fiber plasmonic sensors using 2D molybdenum disulfide nanosheets, while the FOM is relatively low as 
15  RIU$^{-1}$  \cite{wang2018}. 
Also,  a gold nano-disk array coupled to gold mirror is able to achieve high refractive index sensitivity of 853 nm/RIU and a FOM of 126 RIU$^{-1}$ \cite{wang2019}. 
Sensing devices based on grating -assisted plasmonic resonances achieve a high sensitivity 404 nm/RIU and a moderate FOM of 50.3 RIU$^{-1}$ \cite{li2019}. 
Our objective is to couple incident light to SPR is aim to enhance both sensitivity and spectral resolution of plasmonic sensors and hence the FOM. In that sense, very narrow spectral resolutions are reported, being as low as 2.5 nm for thin metallic film nanostructures \cite{Liu2019}, and 0.66 nm for a metamaterial absorber based on an all-metal-grating structure \cite{li2017}.  
When subwavelength high-aspect ratio slits are involved, 
light funnels through these narrow and deep apertures confining the electromagnetic wave within a restricted volume close to them \cite{Bouchon2011,Hughes2016,Li2018}. A metallic grating with subwavelength features  decomposes the field into propagating and evanescent fields, generating a constructive interference close to the aperture \cite{PhysRevLett2011}. 
The spectral response given by this mechanism narrows when the subwavelength aperture is arranged as a multi-layer metallic-dielectric grating structure, instead of a perforated thick metal layer. 
This effect can be used in optical sensing and customized for the 
refractive index of the analyte media.
To improve sensitivity and FOM, the incoming light has to couple efficiently to  surface plasmon resonances (SPR) with the narrowest possible lineshape at the desired wavelength range. 
A wide variety of designs have been proposed to produce these features
\cite{wu2016infrared,wu2017ultra,danilov2018ultra,Liu2019}. 

In this contribution we propose a refractometric sensor based on a multilayer metallic grating embedded in a low-index material ($n_{\rm Cytop}=1.34$, Cytop$^{\rm TM}$ is a trade mark of Asahi Glass Company Chemicals Europe). 
If the structure is terminated by a dielectric surface in contact with the analyte, the subwavelength apertures act as selective transmission filter with a very narrow lineshape. 
When a thin metallic layer covers the whole structure, at some specific wavelenghts light  funnels through the slits of the metallic gratings  towards the sensing surface where SPR are generated.
In both cases, we take advantage of the optimum coupling of the incoming light through light funneling.
This last approach has been already considered by some other authors \cite{Zhu2013}. Here, we invert and refine the design to allow substrate-side illumination and ease the operation of the device.  This arrangement allows spectral interrogation of the device and its coupling to the end of an optical fiber. 
This paper is arranged as follows: In section \ref{sec:model} we explain the geometry of the structure and the results of the simulation in terms of the spectral response and field distributions. 
This geometry has undergone through an optimization process to fix the dimensional parameters of it. Section \ref{sec:deviceanalysis} analyzes the structure as a refractometric sensor, evaluating their performance and checking the range in the index of refraction. Finally, section \ref{sec:conclusions} summarizes the main findings of this paper.

 \section{Modeling and Simulation }

\label{sec:model}

\begin{figure}[h!]
\centering
  \includegraphics[width=0.99\columnwidth]{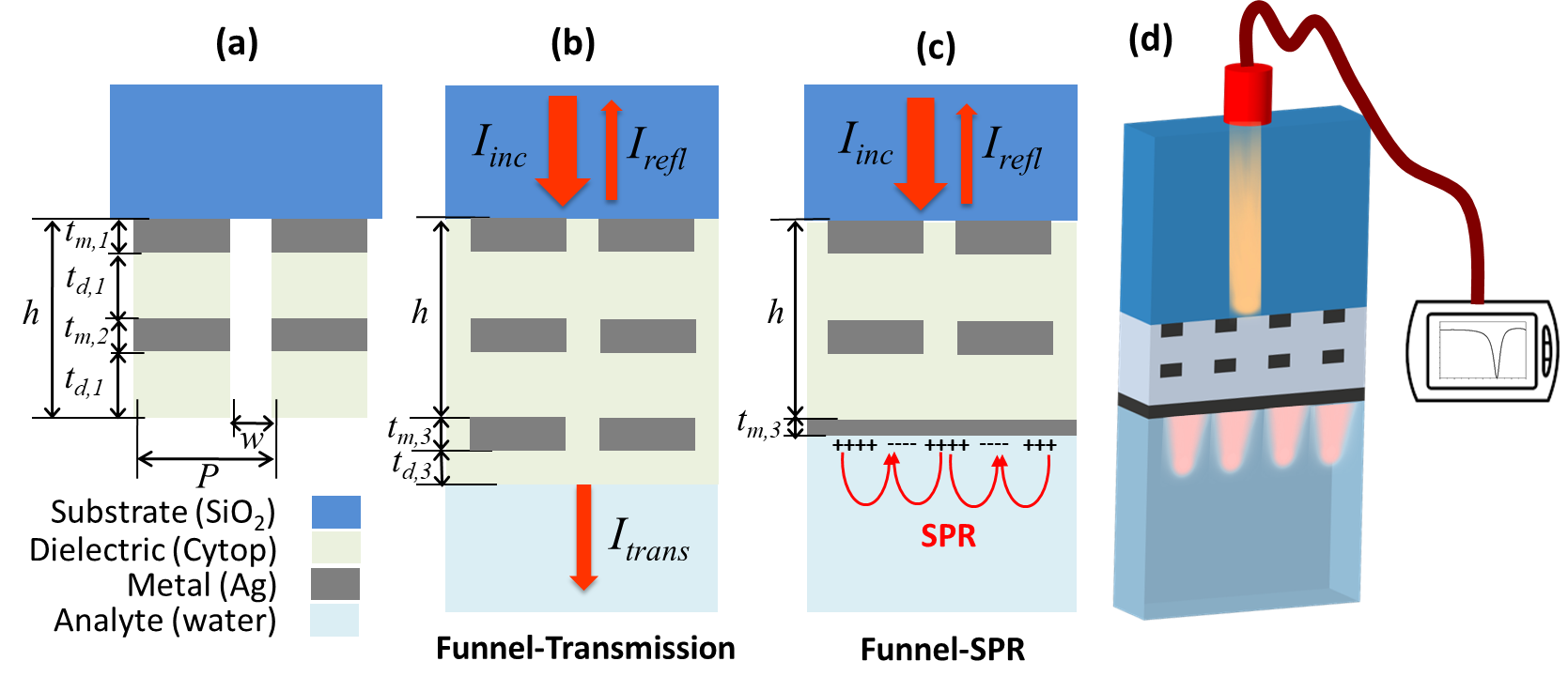}
  \caption{(a) Multi-layer metallic grating structure with periodic narrow slits (period, $P$) of width $w$ and height $h=t_{m,1}+t_{d_1}+t_{m,2}+t_{d,2}$.  (b) The funnel-tranmission design adds another metal ($t_{m,3}$ in thickness) before etching the aperture. Then, the dielectric layer overfills the structure that has a full height of $h+t_{m,3}+t_{d,3}$. The dielectric is in contact with the analyte. 
  The input irradiance, $I_{\rm inc}$, is coming from top, and after interacting with the device is partially absorbed, transmitted ($I_{\rm trans}$), and reflected ($I_{\rm refl}$). These irradiances are represented as red arrows. (c) The funnel-SPR design is terminated with a continuous (no apertures) metallic layer having a thickness of $t_{m,3}$ (this thickness can be different than the one used in (b)). SPRs are generated at the interface between the metal and the analyte. Transmission  towards the analyte is negligible, and the output of the system is read through the reflected irradiance, $I_{\rm refl}$. 
(d) 3D graphical layout of the measurement system for the design in (c). Light is coming from the substrate and the reflectance is also retrieved in the same port, and the nanostructures have been enlarged to show the arrangement.
}
  \label{fig:design}
\end{figure}

The proposed structure is shown in figure \ref{fig:design}. The geometry is described in 2D because the material profile is extruded along the third coordinate.
It is made as a multilayer metallic grating having a narrow deep aperture that reaches the substrate.
The material stack is made of metallic (Ag) and dielectric (Cytop) alternating materials. To allow substrate illumination, we choose SiO$_2$ as the substrate. Each metallic layer has a thickness of $t_{m,1}=t_{m,2}=70$ nm, and the dielectric layer has a thickness of $t_{d,1}=t_{d,2}=150$ nm, that results in a total height of $h=440$ nm.  
The period of the grating is fixed as $P=1150$ nm to spectrally locate  the response around $\lambda_0=1500$ nm, where the availability of light sources and detectors is assured. As it has been previously proved, the grating period can be changed to tune the resonance wavelengths at the desired spectral location: an increase in $P$ produces a red-shift of the spectral response of the structure. The width of the aperture, or slot, is $w= 50$ nm and has been optimized to give the narrowest spectral response and the highest field confinement at the analyte interface (we consider water as the target analyte,  $n_{\rm water}=1.33$) \cite{Elshorbagy2019}.  
This structure can be fabricated by successive depositions of metal and dielectric layers that are selectively etched after exposing the stack with the desired grating pattern through, for example, e-beam lithography. 
Then, the etched trenches are filled with the same low-index dielectric using spin-coating techniques. 

Using this basic arrangement, we analyze two possible variations of the termination of the device. They differ in the material interfacing with the analyte. In the first design, that we name as funnel-transmission design (F-T), we stack a third metallic layer ($t_{m,3}=70$ nm) over the two-layer basic structure (see Fig. \ref{fig:design}.a) before etching the slits. 
Then, after engraving the slits,  the device is finished with a flat dielectric layer (having a thickness $t_{d,3}=70$nm) producing a dielectric/analyte interface (Fig. \ref{fig:design}.b). 
This is possible by overfilling the slots with Cytop$^{\rm TM}$ using a spin-coating fabrication method. 
The second design, named as funnel-SPR design (F-SPR), also fills the slots of the two-metallic stack with the same dielectric materials (Cytop$^{\rm TM}$). 
Then, the structure is terminated with a thin metallic layer (Ag) of $t_{m,3}=45$ nm covering the whole device (this thickness for the third metallic layer is different than the one used in the funnel-tranmission design).  The analyte (water) is in contact with the Ag surface (see Fig. \ref{fig:design}.c). In order to improve biocompatibility and durability, we may add a very thin passivation layer (MgF$_2$)
without compromising the good performance of the device
 \cite{Elshorbagy2019}.

There is a main difference between the physical mechanisms at play in both designs. 
The funnel-transmission device, having a dielectric/analyte interface, is based on the selective transmission of light through the subwavelength apertures towards the analyte. The funnel-SPR design generates SPRs at the metallic/analyte interface. In the following, we use both mechanisms to sense the change in the refractive index of the analyte.

The structure is simulated using the commercial software Comsol Multiphysics (RF module). 
This computational electromagnetism package solves Maxwell equations to obtain the electric and magnetic field distributions, power absorption, reflectance, and transmittance. As far as the model is fully described as a 2D geometry, the calculation can be handled efficiently and optimized using a dedicated desktop workstation. 
The spectral reflectance and transmittance are evaluated using two receiving ports properly set and 
placed on  top and bottom of the structure, respectively.
 Floquet periodic boundary conditions are used on both sides to represent an infinite grating. 
 A plane wave with its magnetic field oriented along the direction of the grating grooves (a TM mode)  is launched from the port located at the substrate side with an amplitude of $H_0=1$ A/m. As a result, the modulus of the calculated magnetic field will be directly considered as field enhancement 
\begin{equation}
{\rm FE}(x,y,z)=\frac{H(x,y,z)}{H_0}
\label{eq:fieldenhancement} 
,\end{equation} 
 where $H(x,y,z)$ is the calculated magnetic field amplitude and $H_0$ is the incident magnetic field amplitude.
The performance of the refractometric sensor is evaluated using the two common parameters of optical sensors: sensitivity, $S_B$, and FOM.

\section{Analysis and results}
\label{sec:deviceanalysis}

\subsection{Optical response and field distribution }
\label{sec:response}

The narrow, high aspect-ratio, aperture of the geometry depicted in figure \ref{fig:design} transmits light selectively within a very narrow band.  The result is the presence of absorption and transmission peaks at those wavelengths where light is able to generate resonances in the structure. 

\begin{figure}[h]
\centering
  \includegraphics[width=1.00\columnwidth]{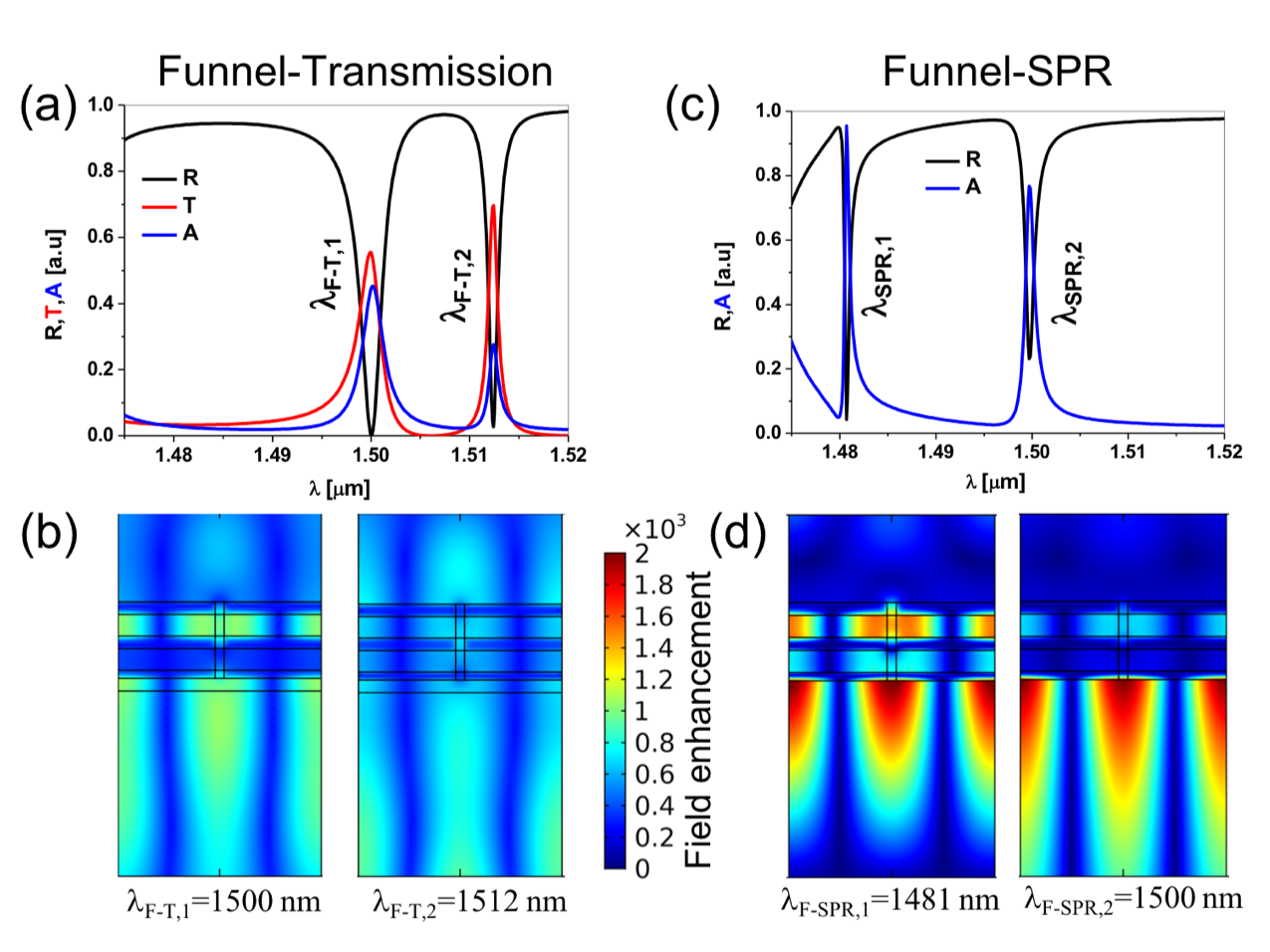}
  \caption{Field enhancement maps (${\rm FE}$, see Eq. (\ref{eq:fieldenhancement})) and spectral reflectance ($R$), transmittance ($T$), and absorption ($A$) for the designs treated in this paper. (a) Spectral behavior for the funnel-transmission design spectral features showing two resonant wavelengths: 
$\lambda_{{\rm F-T},1}=1500$ nm and $\lambda_{{\rm F-T},2}=1512$ nm. (b) Field enhancement maps for these two wavelengths. (c) Spectral behavior for the funnel-SPR design.  (d) Field enhancement maps for the two spectral features of this design at 
$\lambda_{{\rm F-SPR},1}=1481$ nm and $\lambda_{{\rm F-SPR},2}=1500$ nm. In (b) and (d) light comes from top, and the analyte is located at the bottom.
}
  \label{fig:response}
\end{figure}

Both the spectral response and field enhancement maps for the funnel-transmission design are shown in figures \ref{fig:response}.a  and \ref{fig:response}.b.  
In this figure, the incoming light funnels through the structure and is fully coupled  to resonant modes at two wavelengths  located at $\lambda_{{\rm F-T},1}=1500$ nm  and $\lambda_{{\rm F-T},2}=1512$ nm respectively. 
In the first resonance at $\lambda_{{\rm F-T},1}$,  light is strongly confined within the first dielectric grating layer (having a thickness of $t_{d,1}$ in Fig. \ref{fig:design}.a), and coupled to a  plasmonic resonance at the metallic grating close to it. The transmission and absorption are almost equal. For the second resonance, located at $\lambda_{{\rm F-T},2}=1512$ nm,  a weak absorption occurs within the structure, and most of the incoming light is selectively transmitted within a very narrow band. As a result, more field will be available to interact with a large volume of the analyte. The effect of these characteristics on the performance of the structure as a sensor will be presented in the next subsection \ref{sec:sensorperform}.      
    
Figures \ref{fig:response}.c and \ref{fig:response}.d  show the optical response of the funnel-SPR design presented in figure \ref{fig:design}.c. This arrangement has a negligible transmission (the transmitted power does not take into account the evanescent waves generated by the plasmonic resonances). Here, the tiny slots funnel and guide the light towards the metal/dielectric layer where  SPRs are generated. 
The first resonance at $\lambda_{{\rm F-SPR},1} =1481$ nm (see Fig. \ref{fig:response}.c) shows a reflectance close to zero caused by the excitation of SPRs. The distribution of light at each dielectric layer is different, and the first dielectric layer ($t_{m,1}$ in thickness)  hosts a field distribution with larger amplitude. 
The interaction between the SPR resonances and the periodic funneling of light generates an asymmetric lineshape. For the second spectral peak in figure \ref{fig:response}.c ($\lambda_{{\rm F-SPR},2}=1500$ nm), light scatters  from the  grating layers and is partially absorbed by SPR, and partially reflected. Due to this balance between reflection and absorption, the spectral peak (or dip in reflectance) is shallower. The FE map for this wavelength   (see  Fig. \ref{fig:response}.d)  shows a maximum at the analyte location where the surface plasmon wave will interact with the analyte. This is a very interesting feature when considering the device as a refractometric sensor.
Actually, the field enhancement maps (see Figs \ref{fig:response}.b and \ref{fig:response}.d) show how the dielectric of the structure hosts a larger field enhancement for modes 
$\lambda_{{\rm F-T},2}$ and $\lambda_{{\rm F-SPR},2}$  than for modes at 
$\lambda_{{\rm F-T},1}$ and $\lambda_{{\rm F-SPR},1}$.  Therefore, their behaviors when changing the index of refraction of the analyte are different.

From the spectral features shown in figure \ref{fig:response}, we can make some predictions about the performance of the structure when used as refractometric sensors . As far as FOM increases as the resonance is narrower, $\lambda_{{\rm F-T},2}$, and $\lambda_{{\rm F-SPR},1}$ should behave better than the other resonances for the same design. Sensitivity should also follows a similar trend. On the other hand, although this is not analyzed in this contribution, a larger interaction volume, as it happens with $\lambda_{{\rm F-SPR},2}$, should mean a lower limit of detection of the given substance in the analyte.

\subsection{Refractometric sensor performance  }
\label{sec:sensorperform}

The performance of the proposed structures as refractometric sensors is analyzed by calculating the sensitivity, $S_B$, and FOM for different values of the index of refraction of the analyte for both designs and different modes. 
The range in $n_a$ is selected near the index of refraction of water to customize the design for water soluble samples.

\begin{figure}[h]
\centering
  \includegraphics[width=1.0\columnwidth]{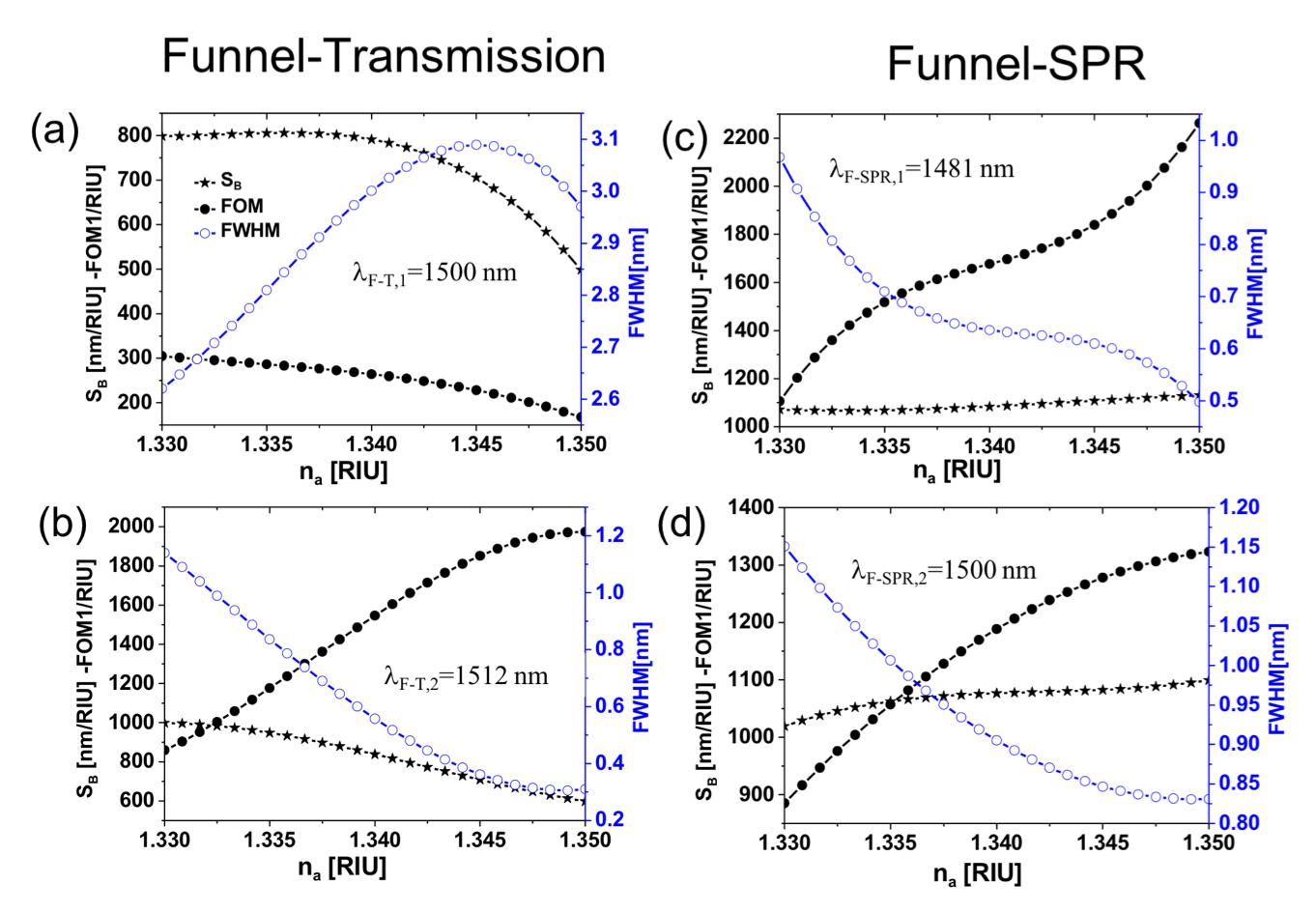}
  \caption{Sensitivity , FOM and FWHM vs. the index of refraction of the analyte for the range $n_a \in [1.33, 1.35]$. The plots for the funnel-transmission design are given in (a) and (b) for wavelengths $\lambda_{{\rm F-T},1}$ and $\lambda_{\rm F-T,2}$ respectively. The responses of the funnel-SPR design are shown in (c) and (d) for $\lambda_{{\rm F-SPR},1}$ and $\lambda_{\rm F-SPR,2}$ respectively.
  }
  \label{fig:sens}
\end{figure}

We first evaluate the performance of the funnel-transmission design (see Fig.  \ref{fig:design}.b). As we saw in the previous section, for the given spectral range, this design  has two spectral features of the reflectance, transmittance, and absorption spectrally located at 
$\lambda_{{\rm F-T},1}$ and $\lambda_{{\rm F-T},2}$ (see Fig. \ref{fig:response}.a).  
The first dip at 1500nm has a strong plasmonic resonance in the embedded metallic grating layers. The results for this dip are shown in figure \ref{fig:sens}.a. The spectral response becomes wider with increasing the analyte refractive index, i.e. FWHM is larger, and as a result, both sensitivity and FOM decrease when increasing $n_a$.
However, we find a range in the index of refraction where $S_B$ remains almost constant. From the definition of the sensitivity (see Eq. (\ref{eq:sens})),  this constancy means a linear behavior of the spectral location of the resonant wavelength, $\lambda_{\rm res}$, with respect to the index of refraction of the analyte, $n_a$. After performing this fitting we conclude that the device could operate linearly at this mode, $\lambda_{{\rm F-T},1}$, in the range $n_a \in [1.33,134]$, following the dependence:  $\lambda_{\rm res}= 805 n_a + 429 $ nm, that results in a linear correlation coefficient $r=0.99998$. The slope of the linear fitting corresponds with the sensitivity value $S_B= 805$ nm/RIU.
These results confirm that the device 
  performs better when the analyte is mainly water, as it happens with blood plasma \cite{Jin2006}. As a consequence of the narrower spectral feature  of the second resonance located at $\lambda_{{\rm F-T},2}= 1512$ nm, the sensor performs better in terms of its sensitivity and FOM. In this mode, both sensitivity and FOM are larger because of the larger transmitted optical energy to the analyte side, and the narrow width, respectively (see the field enhancement maps in Fig. \ref{fig:response}). The FWHM decreases from 2.65 nm for $n_a=1.33$  to 0.3 nm minimum for $n_a=1.35$. That makes the device better  suited for higher values of the refractive index.  

For the funnel-SPR design (see Figs. \ref{fig:design}.c,  \ref{fig:response}.c and \ref{fig:response}.d), we have observed the generation of SPRs at the metal/analyte interface. Their lineshapes are now dependent on the coupling of the periodic funneling caused by the grating, and the SPR. 
This design  shows high sensitivity and ultra-high FOM values reaching 1130 nm/RIU and 2100 RIU$^{-1}$, respectively for the mode located at $\lambda_{{\rm F-SPR,1}}=1481$ nm. This behavior is presented in Fig. \ref{fig:sens}.c in terms of the index of refraction of the analyte.
For the second resonance  at $\lambda_{{\rm F-SPR},2}=1500$ nm, only one layer of the grating scatters the funneled light toward the metal film, so the coupling is weaker compared to $\lambda_{{\rm F-SPR},1}$.  The complete results of the resonance  at $\lambda_{{\rm F-SPR},2}$ are shown in figure \ref{fig:sens}.d.

A summary of the maximum values of sensitivity and FOM for both designs and modes are listed in table \ref{tab:sum}. 
 
\begin{table}[h!]
\caption{ Maximum values of sensitivity and FOM. \label{tab:sum}}
\centering
\begin{tabular}{rccc}
\toprule
Design & $\lambda$ (nm)  & $S_B$ ($@ n_a$) (nm/RIU) & FOM ($@ n_a$) (1/RIU) \\  
\hline \hline
Funnel-Transmission & $\lambda_{{\rm F-T},1}=1500$ & 800 ($@$ 1.33)  & 300 ($@$ 1.33) \\
					& $\lambda_{{\rm F-T},1}=1512$ & 1000 ($@$ 1.33)  & 1970 ($@$ 1.35) \\
					\hline
Funnel-SPR & $\lambda_{{\rm F-SPR},1}=1481$ & 1130 ($@$ 1.35) & 2100 ($@$ 1.35) \\
			& $\lambda_{{\rm F-SPR},2}=1500$ & 1100 ($@$ 1.35) & 1330 ($@$ 1.35) \\
\bottomrule
\end{tabular}
\end{table}

\section{Conclusions}
\label{sec:conclusions}

Subwavelength deep apertures in metallic surfaces funnel light and allow high field enhancement for selected wavelengths. 
We have analyzed a metallic-dielectric multilayer structure where high-aspect ratio apertures makes possible funneling. The geometry and material arrangement is prepared for refractometric sensing of water-solvent analytes. 
Two designs are proposed that differ in the termination of the device.
One of them adds a third grating and is terminated by a dielectric layer (funnel-transmission design).
The other has a metallic thin layer in contact with the analyte (funnel-SPR design). 
The proposed devices can be realized using nanofabrication tools. 
The metallic and dielectric materials at the multilayer stack can be deposited by thermal evaporation, after etching the apertures, the dielectric can be spin coated on the structure to  fill the trenches. 
The termination of the device only require conventional spin-coating and/or thermal deposition. Once fabricated, this system can be interrogated spectrally and integrated for normal incidence conditions (for example, at the tip of an optical fiber).

The system is optimized to generate high field enhancement and very narrow spectral features. A high value of the field improves the interaction with the analyte, and a narrower spectral response produces larger values of the sensitivity and FOM of the device. 
The analysis of the proposed designs is completed by evaluating their capabilities as refractometric sensors. The maximum values of both $S_B$ and FOM are obtained for the funnel-SPR design reaching values of 1130 nm/RIU and 2100 RIU$^{-1}$ respectively. These values can be considered very competitive with existing proposals.

\section*{Acknowledgements}
This work was partially supported by the Egyptian Ministry of Higher Education missions section.


%%%REFERENCES%%%
%\bibliographystyle{spmpsci} 
%\bibliographystyle{IEEEtran}
%\bibliography{funnelsens} 


\end{document}